\documentclass[showpacs, twocolumn]{revtex4-1}
\usepackage{graphicx}
\usepackage{amsmath}
\usepackage{appendix}

\begin{document}

\title{Moving binary Bose-Einstein condensates in a weak random potential}

\author{Abdel\^{a}ali Boudjem\^{a}a}
\affiliation{Department of Physics,  Faculty of Exact Sciences and Informatics, Hassiba Benbouali University of Chlef P.O. Box 78, 02000, Ouled Fares, Chlef, Algeria.}
\email {a.boudjemaa@univ-chlef.dz}


\begin{abstract}

We study the behavior of moving Bose-Bose mixtures in a weak disordered potential in the realm of the Bogoliubov-Huang-Meng theory.
Corrections due to the quantum fluctuations, disorder effects and the relative motion of two fluids to the glassy fraction, the condensed depletion, the anomalous density,
and the equation of state of each species are obtained analytically for small velocity.
We show that the intriguing interplay of the relative motion and the disorder potential could not only change the stability condition, 
but destroy also the localization process in the two condensates preventing the formation of a Bose glass state. 
Unexpectedly, we find that the quantum fluctuations reduce with the velocity of the two fluids.
The obtained theoretical predictions are checked by our numerical results.

\end{abstract}


\maketitle


Recently, quantum Bose mixtures have afforded a fascinating glimpse into the cold atoms world since they display rich phase separation behavior. 
Mixtures of two species, or two isotopes of a single species, or two-component Bose-Einstein condensates (BECs) were all successfully realized experimentally \cite{Mya,Hall,Mad,Cab,Sem,Pap,Sug,Mog, McC,Ler,Pasq,Wack,Wang,Igor}.
One of the most elusive feature of such ensembles is their miscibility.
At zero temperature, the miscibility-immiscibility  phase transition  depends on the ratio of the intra- and interspecies interactions via  
the miscibility parameter $\Delta=g_1 g_2/g_{12}^2$, where $(g_1, g_2)$ and $g_{12}$ are the intraspecies and interspecies scattering lengths, respectively.
For $\Delta >1$, the two condensates overlap in space while they phase separate when $\Delta <1$.
However, at finite temperature, deviation from this condition has been very recently predicted in Refs.\cite{Boudj,Boudj1,Ota}. 

It  has been found that binary BECs \cite {Hall,Boudj,Sinatra,Lee} have complex motion that tended to preserve the total density 
but quickly damped to a stationary state with non-negligible component overlap.
These moving wave packets of BECs have the possibility to collide and move through each other \cite{Hall, Koz, Hag, Mad, Band}.
The influence of the relative motion of two species on the stability condition of the system has been analyzed in Ref.\cite{Yuk}.

On the other hand, Bose-Bose mixtures in random potentials have attracted immense attention in recent years \cite{Wehr, Nied, Xi,Mard,BB}.
Quite recently, we have found that the competition between the interspecies interactions and the disorder potential
may affect the localization, the stability condition, the thermodynamics, and the superfluid density in each component of the mixture \cite{BB}.

In this letter, we tackle the problem of a dirty moving symmetric Bose mixture gas using the Bogoliubov-Huang-Meng theory \cite{HM}.
Such a system can shed a new light on the quantum transport of collective excitations due 
to an unprecedented control of interspecies interactions and the possibilities of conception controlled random. 
Moreover, disordered Bose mixtures with moving components may offer also a unique opportunity to study  Anderson localization.
They also play an important role in the fundamental understanding of the intra- and interspecies interactions effects on the diffusion of expanding condensates.

The aim of this work is to study the impacts of a weak disorder potential with delta-correlated disorder on the properties of moving Bose mixtures. 
The velocity of each component is assumed to be considerably less than the speed of sound in the mixture liquid. 
Based on the Bogoliubov theory, analytical formulas for the relevant physical quantities such as the glassy density, the quantum depletion, the anomalous density, and
the equation of state (EoS) of each component are derived. Condition under which the theory works is properly established.
Our results reveal that the relative motion of  two BECs may reduce the localization of atoms and the quantum fluctuations in each species.
This leads to prohibit the formation of the glass state and hence, allows the emergence of weakly correlated and ultradilute BECs.
It is shown that the interplay between inter and intra-species interaction tends to increase the glassy fraction and the condensed depletion, while it lowers the anomalous density.
We demonstrate in addition that the EoS rises with both the velocity and the interactions.
Our numerical calculations confirm this scenario for parameters relevant to the recent experiments of Ref.\cite{Cab}.



Consider two-component BECs  with equal masses, $m_1=m_2=m$, in a weak disorder potential $U({\bf r})$, labeling the components by the index $j$.
A quantitative condition for weak disorder will be determined later on.
Bogoliubov quasiparticles and disorder-induced fluctuations decouple in the lowest order \cite{HM}, 
implying that at zero temperature and in the dilute regime, the dynamics of such a system can be governed by the coupled Gross-Pitaevskii equations (GPE) \cite{BB}
\begin{align} \label{GP}
	i\hbar \frac{\partial \phi_j({\bf r},t)}{\partial t}&= \bigg (\frac{-\hbar^2}{2m}\nabla^2+U({\bf r})+g_{j}|\phi_j({\bf r},t)|^2 \\
&+g_{12}|\phi_{3-j}({\bf r},t)|^2\bigg)\phi_j({\bf r},t), \nonumber
\end{align}	
where the intra- and the interspecies coupling constants are given, respectively in terms of
the intra- and interspecies $s$-wave scattering length by $g_j=4\pi\hbar^2 a_j/m_j$ and $g_{12}=2\pi\hbar^2a_{12}/m$.
The disorder potential  should fulfil the following statistical properties
$\langle U(\mathbf r)\rangle=0$, and  $\langle U(\mathbf r) U(\mathbf r')\rangle=R (\mathbf r,\mathbf r')$,
where $ \langle \bullet \rangle$ denotes the disorder ensemble average and $R (\mathbf r,\mathbf r')$ is the disorder correlation function. 

Now we assume a mixture of moving components, where each component moves with the same velocity $v_1 =v_2 =v$.
The motion of the mixture is done by means of the Galilean transformation:
\begin{equation}\label{GP1}
	\phi({\bf r},t)=\phi({\bf r},t)\exp{ \left(\frac{i}{\hbar} m_j {\bf v}.{\bf r}\right)}.
\end{equation}
Introducing Eq.(\ref{GP1}) into Eq.(\ref{GP}), the above coupled GPE can be rewritten as follows:
\begin{align}\label{GP2}
	i\hbar \frac{\partial \phi_j({\bf r},t)}{\partial t}&=\bigg(\frac{p^2}{2m_j} +{\bf p}.{\bf v}+\frac{1}{ 2}m_j v^2+U({\bf r}) \\
&+g_j|\phi_j({\bf r},t)|^2+g_{12}|\phi_{3-j}({\bf r},t)|^2\bigg)\phi_j({\bf r},t),\nonumber
\end{align}
where ${\bf p}= -i \hbar {\bf \nabla}$ is the kinetic-energy operator.


The calculation of the Bogoliubov spectrum and quantum fluctuations for moving Bose mixtures amounts to solving the Bogoluibov-de-Gennes equations.
To do so, we linearize Eqs.(\ref{GP2}) around static solutions $\phi_{0j}$ using the transformation:
\begin{equation} \label{BogTran}
	\phi_j({\bf r},t)=\left[\phi_{0j} ({\bf r})+\delta\phi_j({\bf r},t) \right]\exp{\left(- i\mu_j t/\hbar \right)},
\end{equation}
where $\mu_j$ are chemical potentials related with bosonic components, and $\delta\phi_j({\bf r},t) \ll \phi_{0j}$ are small quantum fluctuations.
Let us consider the case of a uniform mixture and focus on the situation where the external potential is very weak, hence 
$U({\bf r})$ describes the local spatial fluctuations around the homogeneous background \cite{Gaul}.
Substituting Eq.(\ref{BogTran}) into Eq.(\ref{GP2}) and keep in mind that the wavefunctions are real-valued 
($\phi_{0j}=\phi_{0j}^*=\sqrt{n_j}$, and $\phi_{0\,3-j}=\phi_{0\,3-j}^*=\sqrt{ n_{3-j}}$), we obtain up to zeroth-order 
\begin{align} 
	\mu_j\sqrt{n_j}&=\bigg[\frac{p^2}{2m_j} +{\bf p}.{\bf v}+\frac{1}{ 2}m_j v^2+U+g_j n_j \\
&+g_{12} n_{3-j} \bigg] \sqrt{n_j}. \nonumber
\end{align}
The first-order terms yield
\begin{align}  \label{GP3}
	i\hbar\frac{\partial\delta\phi_j({\bf r},t)}{\partial t}&=\bigg(\frac{p^2}{2m_j} +{\bf p}.{\bf v}+\frac{1}{ 2} m_j v^2+U+2g_j n_j \\
             &+g_{12} n_{3-j}  -\mu\bigg)\delta\phi_j({\bf r},t) +g_j n_j\delta\phi_j^*({\bf r},t)\nonumber\\
	&+g_{12}\sqrt{n_j n_{3-j}} \left[\delta\phi_{3-j}^*({\bf r},t)+\delta\phi_{3-j} ({\bf r},t)\right].\nonumber
\end{align}
To calculate the energy of elementary excitations, we introduce transformation: 
\begin{align} \label{HM1}
	\delta\phi_j({\bf r},t)&=u_{jp}\exp({i{\bf p.r}/\hbar-i\varepsilon_p t/\hbar}) \\
&+v_{jp}\exp({i{\bf p. r}/\hbar+i\varepsilon_p t/\hbar})-\beta_{jp},\nonumber
\end{align}
where $u_{jp}$ and $v_{jp}$ are the Bogoliubov amplitudes, $\beta_{jp}=\sqrt{n_j/V} \left(|u_{jp}+v_{jp}|^2/\varepsilon_p\right) U_p$ \cite {Yuk}, 
where $U_p$ is the Fourier transform of the external potential $U({\bf r})$ and $V$ denotes the volume.
For two decoupled BECs, Eq.(\ref{HM1}) reduces to the standard Huang-Meng transformation \cite{HM}.
Thus, the chemical potential takes the form:
\begin{equation} \label{Chim}
	\mu_j=\frac{1}{2}m_jv^2+U+g_j n_j +g_{12} n_{3-j}.
\end{equation}
Inserting Eqs.(\ref{Chim}) and (\ref{HM1}) into Eq.(\ref{GP3}), we obtain the coupled Bogoluibov-de-Gennes equations for 
a mixture with moving species:
\begin{align} \label{BdG1}
	\varepsilon_p u_{jp}&=\left(E_p+ {\bf p}.{\bf v}+g_j|\phi_j|^2\right) u_{jp}+g_jn_jv_{jp}\\
&+g_{12}\sqrt{n_jn_{3-j}}(u_{jp}+v_{jp}), \nonumber
\end{align}
and 
\begin{align} \label{BdG2}
	-\varepsilon_p v_{jp}&=\left(E_p+{\bf p}.{\bf v}+g_j|\phi_j|^2\right) v_{jp}+g_jn_ju_{jp} \\
&+g_{12}\sqrt{n_jn_{3-j}}(u_{jp}+v_{jp}),  \nonumber
\end{align}
where $ E_p= p^2/ 2m$ is the free particle energy.

From now on we assume a symmetric mixture with $n_1=n_2=n$ and $g_1=g_2=g$.
A straightforward calculation leads to the following expressions for the quasiparticle amplitudes $u_{p\pm}, v_{p\pm}$:
\begin{equation}
	u_{p\pm},v_{p\pm} =\frac{ 1}{ 2} \left (\sqrt{ \frac{\varepsilon_{p\pm} } {E_p+p v \cos\theta} }\pm\sqrt{ \frac{E_p+p v \cos\theta} {\varepsilon_{p\pm}} } \right),
\end{equation}
and the gapless excitations spectrum of the system reads
\begin{equation} \label{Bog}
	\varepsilon_{p\pm}=\sqrt{\left(E_p +p v \cos\theta\right )^2+2 m c_{s\pm}^2 \left(E_p+p v \cos \theta\right) },
\end{equation}
where $c_{s+}= c_s\sqrt{\delta g_{+}/g}$, and $c_{s-}= c_s \sqrt{\delta g_{-}/g}$ are the sound velocities in the density and spin channels \cite{Ota19}, respectively,
with $c_s=\sqrt{gn/m}$ being the usual sound velocity of a single BEC, $\delta g_{\pm}=g(1\pm g_{12})$, and $\theta$ is the angle between the two vectors ${\bf p}$ and  ${\bf v}$. 
For $ v=0$, one reproduces the standard density and spin excitations for immovable Bose mixtures: $\varepsilon_{p \pm}=p\sqrt{(p/2m)^2+c_{s\pm}}$ \cite{Boudj2}.
The stability of the spectrum (\ref{Bog}) requires the conditions:  $\delta g_{\pm} >0 $ and $\theta=2 \ell \pi$, with $\ell$ being an integer or equivalently,
$\delta g_{\pm} < 0 $ and $\theta=(2\ell+1)\pi$.
Note that the stability condition could also be modified as a result of the relative motion of two liquids \cite{Yuk}. 
For quantum droplets with $\delta g_+<0$, the $\varepsilon_{p +}$ density energy is imaginary. 
In this regime, the mean-field theory would result in energy $\propto n^2$ leading to a collapse of a homogeneous state.
The beyond mean-field Lee-Huang-Yang (LHY) corrections which provide a repulsive term $\propto n^{5/2}$ eliminates the mechanical instability 
(see e.g. \cite{Petrov,Boudj2} and references therein).

Similarly to the single disordered BEC, the spectrum (\ref{Bog}) is independent of the disorder potential.
This independence justifies the neglect of the noncondensed and anomalous fluctuations in the Bogoliubov quasiparticle spectrum.
Such an assumption is necessary to guarantee that the excitations spectrum  has no energy gap \cite{HM, Mich}.
Obviously, if one takes into consideration higher-order quantum fluctuations in the spectrum \cite{Gaul,Boudj,Boudj1,Yuk}, it would depend on the random potential.


\begin{figure}
	\includegraphics[width=0.46 \linewidth]{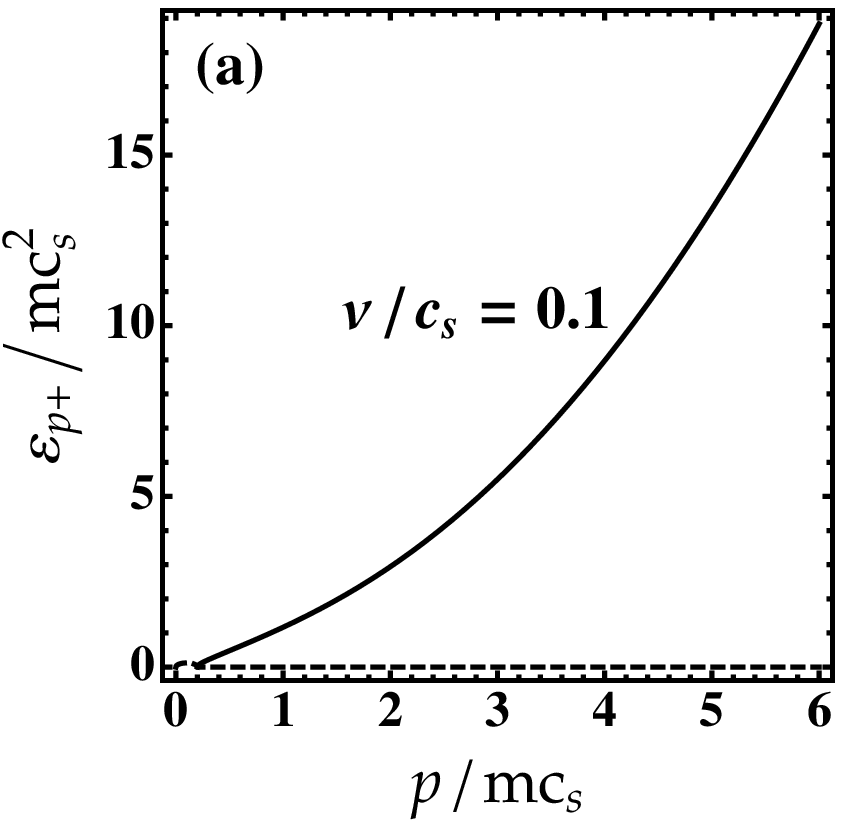}
        \includegraphics[width=0.45 \linewidth]{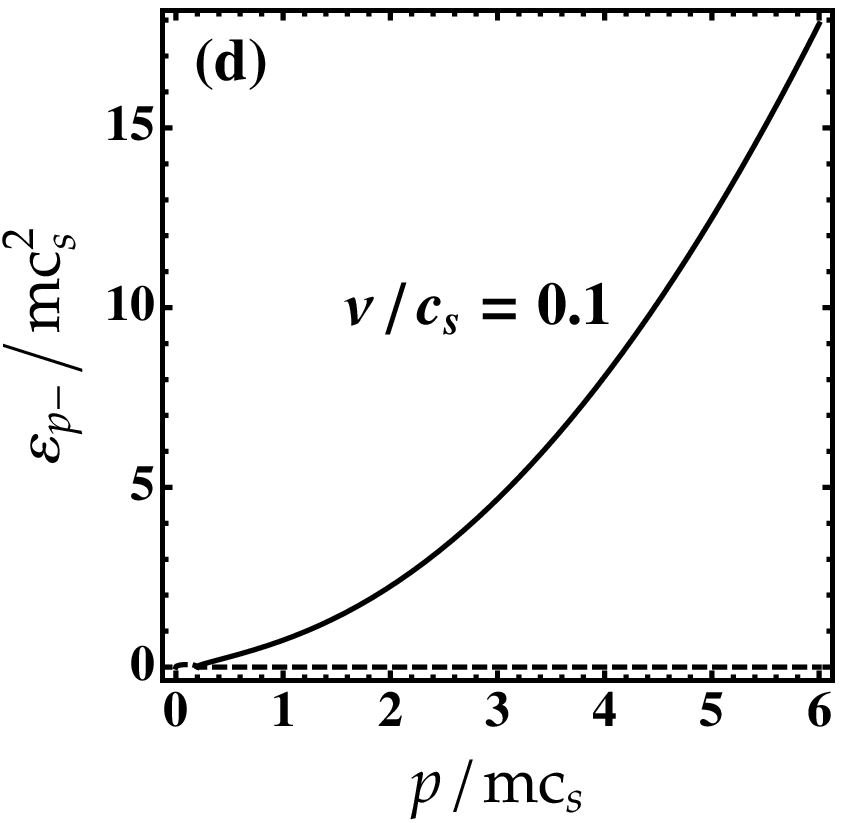}
	\includegraphics[width=0.47 \linewidth]{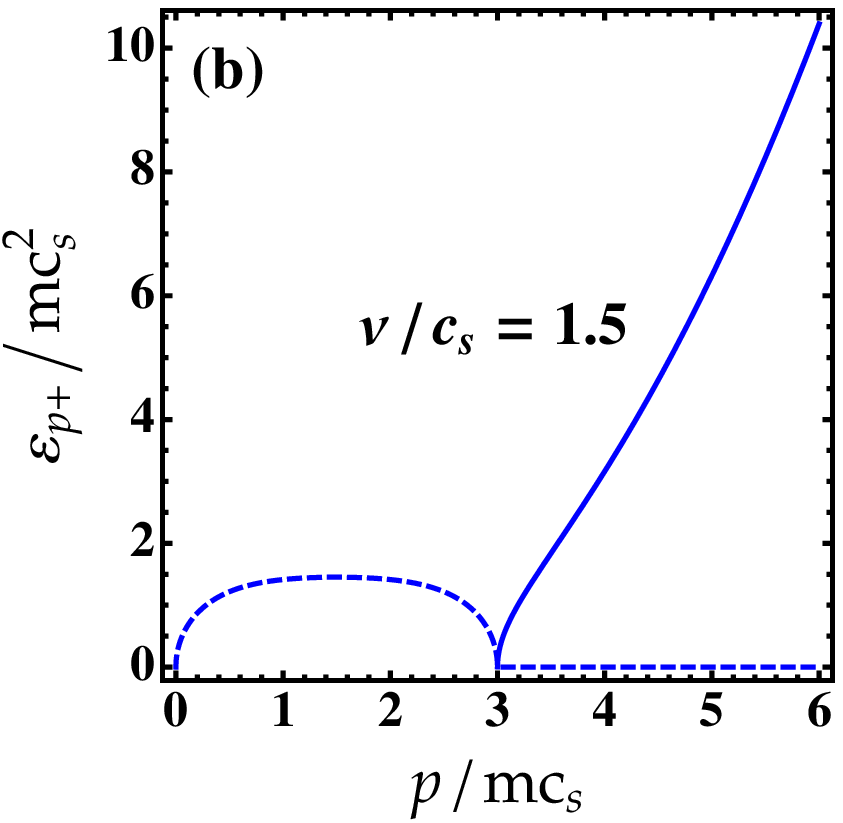}
	\includegraphics[width=0.45 \linewidth]{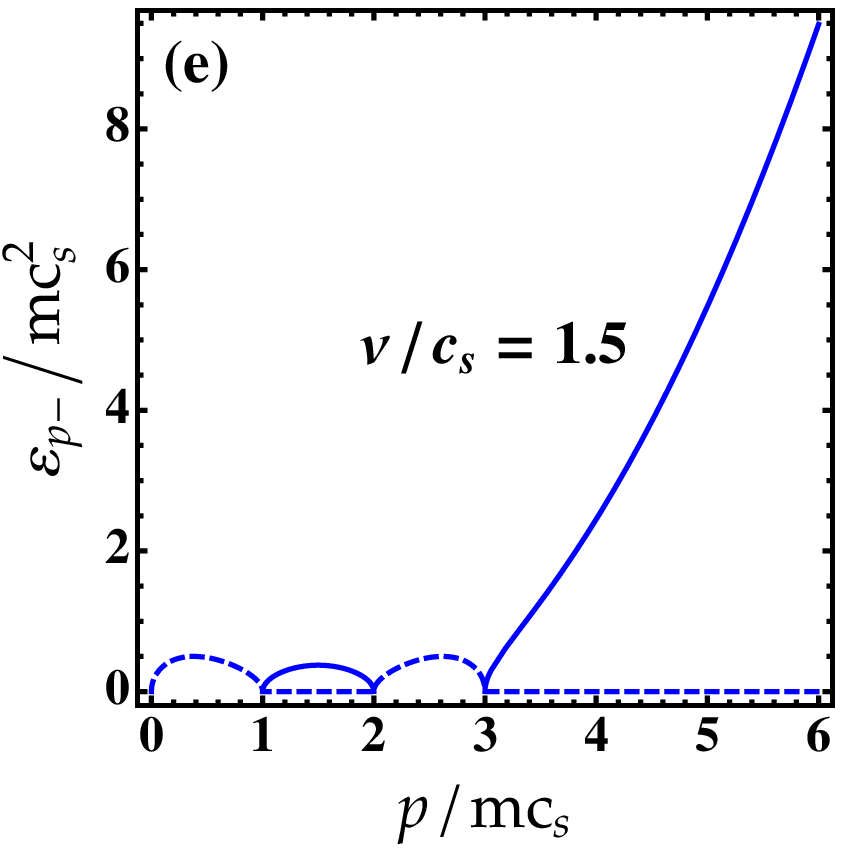}
	\includegraphics[width=0.46 \linewidth]{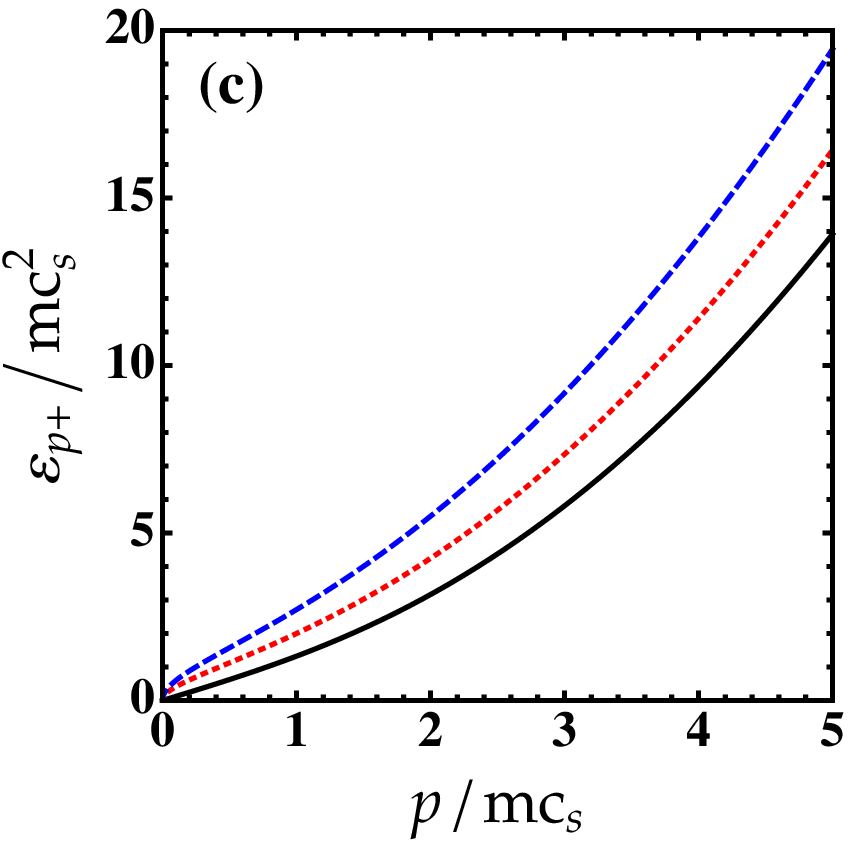}
	\includegraphics[width=0.46 \linewidth]{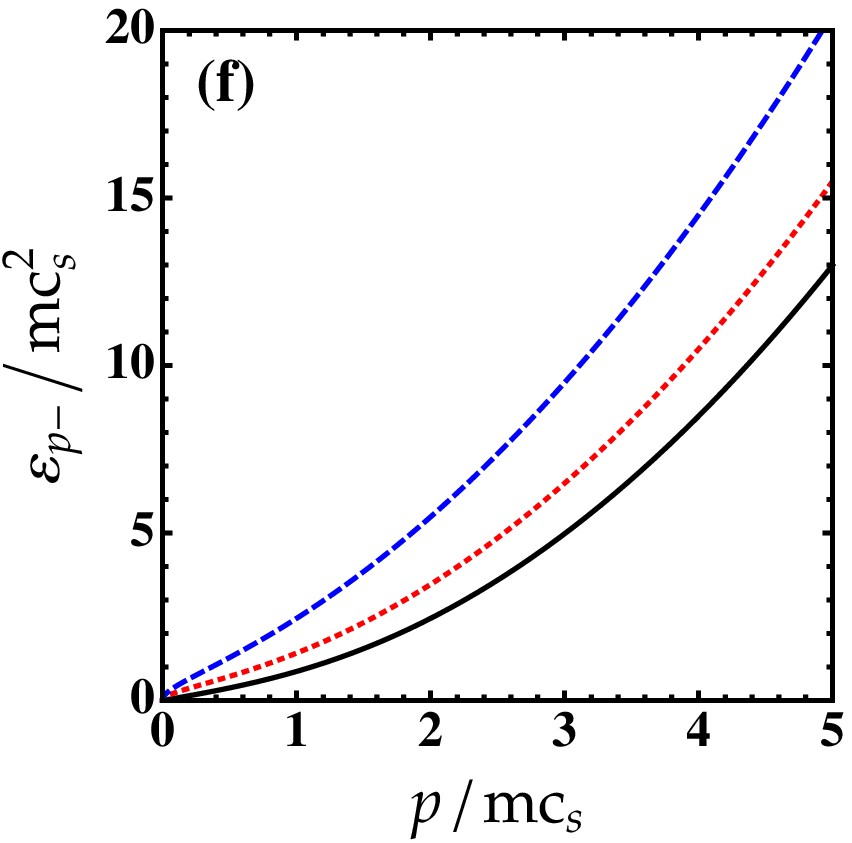}
	\caption{(a) Bogoliubov excitations energy $\varepsilon_{p+}$ from Eq.(\ref{Bog}), for $\theta=\pi$, $a_{12}/a=0.5$ and $v/c_s=0.5$.
 Solid lines correspond to the real part of the spectrum. Dashed lines correspond to the imaginary part of the spectrum.
(b) The same as Fig.\ref{sepBog}.(a) but for $v/c_s=1.5$.
(c) Bogoliubov excitations energy $\varepsilon_{p+}$ for $\theta=2\pi$, $a_{12}/a=0.5$ and several values of $v/c_s$. 
Black line: $v/c_s=0$. Red dotted line: $v/c_s=0.5$. Blue dashed line: $v/c_s=1.5$.
Figs.\ref{sepBog}.(d)-(f)  are the same as Figs.\ref{sepBog}.(a)-(c) but for $\varepsilon_{p-}$.}
\label {sepBog}
\end{figure}

In Fig.\ref{sepBog} we plot the spectrum (\ref{Bog}) for different values of the relative velocity and the angle $\theta$.
The dispersion  $\varepsilon_{p\pm}$ has an imaginary part for $\theta=\pi$ regardless of the value of the velocity and the strength of the interspecies interaction.
The spatially modulated perturbations grow with ratio $v/c_s$ as is seen in Figs.\ref{sepBog}.(a), (b), (d) and (e).
The fastest growth occurs for the wave number $p_{max}=\partial \varepsilon_{p\pm}/ \partial p=0$ that gives a maximum of $\varepsilon_{p\pm}$.
One should stress that the interspecies interaction $a_{12}/a$ may  also affect the  modulation of $\varepsilon_{p\pm}$.
Figures \ref{sepBog}.(c) and (f) show that for $\theta=2\pi$, the Bogoliubov spectrum $\varepsilon_{p\pm}=\sqrt{\left(E_p +p v \right )^2+2 m c_{s\pm}^2 \left(E_p+p v \right)}$ 
is stable in the whole range of the momentum $p$. 
We observe also that the relative velocity $v/c_s$ rises the spectrum $\varepsilon_{p\pm}$ leading to diminish the density of noncondensed atoms and 
the anomalous density (see below).


We assume henceforth that $\delta g_{\pm}>0$ and $\theta=2 \ell \pi$. 
In this case, the minimum value possible for $v$ which must be kept positive,
can be obtained in lowest value of $p$ i.e in the phonon regime: $v_c=c_s \sqrt{\delta g_{\pm}/g}$, below which there is no solution for $v$.

At zero temperature, the condensate depletion and the anomalous density
for each component are defined, respectively as \cite{Boudj5}: $\tilde{n}_{\pm}=V^{-1} \sum_{\bf p} [v_{p \pm}^2+ \langle |\beta_p|^2 \rangle]$,
and $\tilde{m}_{\pm}=V^{-1} \sum_{\bf p} [-u_{p \pm}v_{p \pm} +\langle |\beta_p|^2 \rangle]$.
Working in the thermodynamic limit, the sum over $p$ can be replaced  by the integral $\sum_{\bf p} \rightarrow V\int_{ 0}^{ \infty}  d{\bf p}/(2\pi\hbar)^3 $.
Thus, the noncondensed and anomalous densities take the form :
\begin{align}\label{Dep} 
\tilde n_{\pm}&=\tilde n_{\pm}^0+n_{R\pm} \\
&=\int \frac{p^2d p}{4 \pi^2 \hbar^3} \left[\frac{E_p+p v + mc_{s\pm}^2} {\varepsilon_{p \pm}}-1\right]+n_{R\pm},\nonumber
\end{align}
and
\begin{equation}\label{mDep} 
\tilde m_{\pm}=\tilde m_{\pm}^0+n_{R\pm}=-\int \frac{p^2d p}{4 \pi^2 \hbar^3} \frac{mc_{s\pm}^2 } {\varepsilon_{p \pm}}+n_{R\pm},
\end{equation}
where the glassy fraction $n_{R\pm}$ reads      
\begin{equation} \label{glass}
n_{R\pm}=\frac{ 1}{ V}\sum_p \langle |\beta_{p\pm}|^2 \rangle=n\int_{ 0}^{ \infty}\frac{ p^2dp}{ 4\pi^2\hbar^3}\frac{ R_p(E_p+p v )^2}{ \varepsilon_{p\pm}^4},
\end{equation}
where $R_p=\langle |U_{p}|^2 \rangle$.
The first term in Eqs.(\ref{Dep}) and (\ref{mDep}) originates from the quantum fluctuations.
Whereas, $n_R$ defined in Eq.(\ref{glass}) accounts for the disorder fluctuation known also as {\it glassy fraction} is analog to the Edwards-Anderson order parameter 
of a spin glass \cite{Edw,Yuk1}. It arises from the accumulation of density near the potential minima and density depletion around the maxima. 
The emergence of glassy fraction in Bose systems does not mean that  the whole system is transformed into into the Bose glass phase \cite{Yuk1}.
Note that the validity of the present approach requires the inequality: $\tilde n \ll n$ \cite{HM,Axel,Boudj3}.


In quantum Bose gases, the EoS is important since it characterizes the thermodynamic properties of the system.
Corrections to the EoS owing to the Lee-Huang-Yang (LHY) quantum fluctuations for Bose mixtures can be given via the relation \cite{Boudj,Boudj2}:
\begin{equation}\label {chim}
\mu_{\text{LHY} \pm} = \frac{g}{V}\sum_p [v_{p \pm}^2- u_{p \pm} v_{p \pm}]= g (\tilde n_{\pm}+\tilde m_{\pm}), 
\end{equation}
where $\tilde n_{\pm}$ and $\tilde m_{\pm}$ are defined in Eqs.(\ref{Dep}) and (\ref{mDep}).


To be concrete,  we consider in what follows the white noise random potential, which assumes a delta distribution  
\begin{equation}  \label{delt}
R ({\bf r}- {\bf r'})=R_0\delta ({\bf r}- {\bf r'}), 
\end{equation}
where $R_0$ is the disorder strength with dimension (energy)$^2$ $\times$ (length)$^3$. 
The model (\ref{delt}) is valid when the correlation length of the correlation function $R ({\bf r}- {\bf r'})$ is sufficiently shorter than the healing length
(see e.g. \cite{Yuk1,Axel,Boudj3,Boudj4}). 


Integrals (\ref{Dep})-(\ref{glass}) are infinite for the Bogoliubov energy (\ref{Bog}).
Assuming that the velocity of the mixture is small ($v \ll v_c=c_s \sqrt{\delta g_{\pm}/g}$) we expand the above integrals in powers of $v$ up to first-order.
The zero-order terms give exactly the usual expressions for the condensate depletion, the anomalous density, and the EoS of immovable Bose mixtures.

After a straightforward calculation, we get for the glassy fraction:
\begin{equation} \label{glass1}
n_{R\pm} \simeq n_{\text{HM}\pm} \bigg (1- \frac{2 v} {\pi c_{\pm}} \bigg),
\end{equation}
where $n_{\text{HM}\pm}= n mR_0/(4\pi \hbar^3 {c_{s\pm}})$.
For $v=0$,  one can reproduce the result of immovable dirty binary BECs obtained in our recent work \cite{BB}. 
Equation (\ref{glass1}) shows that $n_{R\pm}$ is decreasing with increasing $v /c_{\pm}$ indicating that the velocity of two BECs
lead to tune the disorder fluctuations ensuring the existence of the condensate.

Working out the quantum contribution following the same procedure, the condensate depletion of each species takes the form:
\begin{equation}\label{Dep1} 
\tilde n_{\pm}^0 \simeq \frac{m^3c_{s\pm}^3}{3 \pi^2 \hbar^3} \bigg(1-\frac{3v}{2c_{\pm}}\bigg) . 
\end{equation}
Again the increase in velocity of two fluids in turn reduces the density of noncondensed atoms $\tilde n_{\pm}^0$ and hence, augments the density of condensed atoms.
One may argue that noncondensed particles screen the motion of atoms through the system.


\begin{figure}
	\includegraphics[width=0.8 \linewidth]{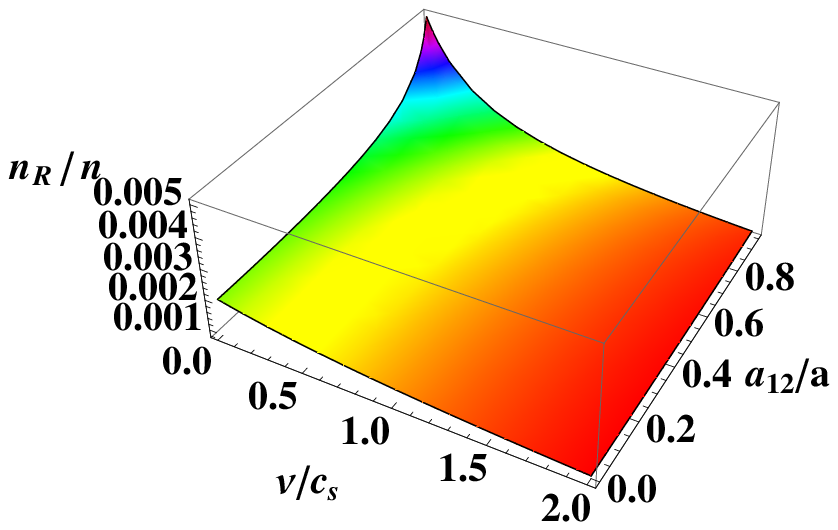}
	\includegraphics[width=0.8 \linewidth]{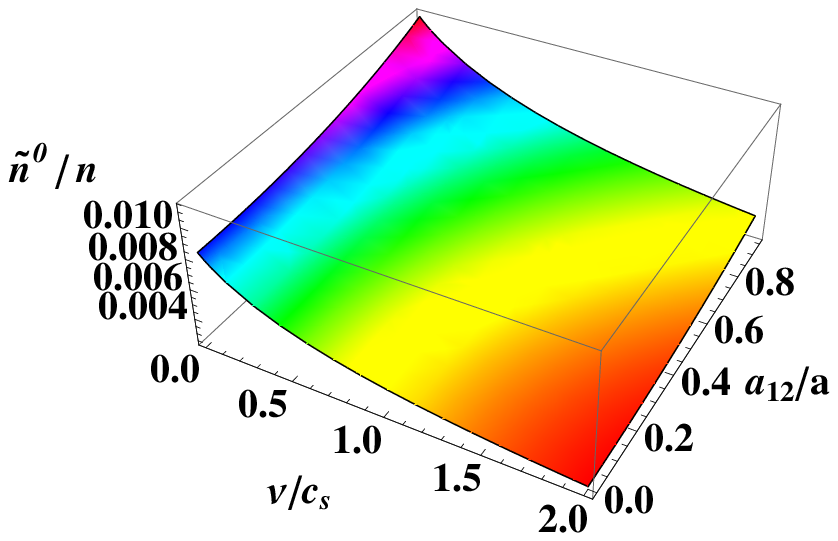}
	\includegraphics[width=0.8 \linewidth]{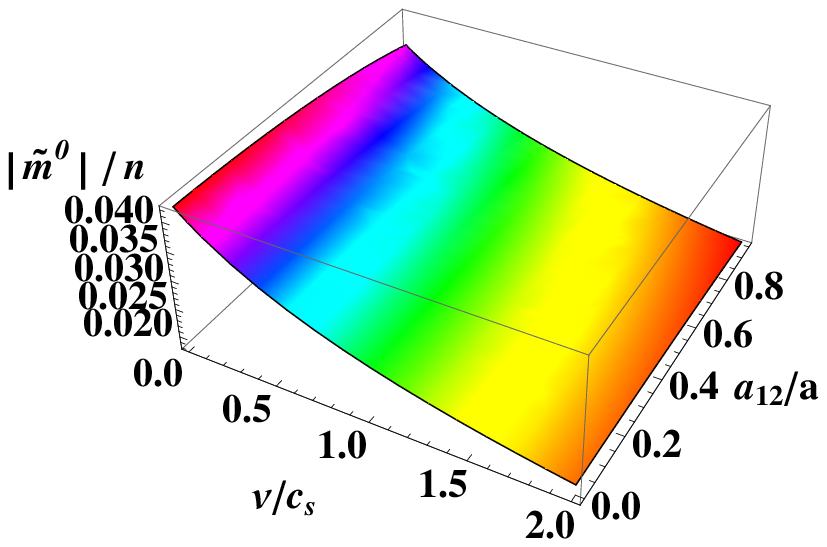}
	\caption{(Top) Glassy fraction from Eq.(\ref{glass}). (Middle) Noncondensed fraction of a cleaned Bose mixture from Eq.(\ref{Dep}). 
        (Bottom) Anomalous fraction of a cleaned Bose mixture from Eq.(\ref{mDep}) as a function of the mixture velocity $v/c_s$ and the interspecies interaction $a_{12}/a$.
Parameters are : $R'=0.05$, and $na^3\sim 10^{-5}$ \cite{Cab}.}
\label {AND}
\end{figure}

Corrections to the anomalous density due to the relative motion contribution is the same as for the noncondensed density.
The only difference comes from the quantum term which diverges at high momentum.
This ultraviolet divergence is well-known  and arises from the contact potential.  
One strategy to overvome this problem is the dimensional regularization which is valid for weak interacions \cite{Yuk, Boudj}. 
After regularization we obtain for the anomalous density $\tilde m_{\pm}^0 \simeq(mc_{s\pm})^3/ \pi^2 \hbar^3 \big(1-3v/2c_{\pm}\big)$, 
which is larger than $\tilde n_{\pm}^0$.


Collecting all terms together, we obtain useful expressions for the total depletion $\tilde n=\sum_{\pm}\tilde n_{\pm}$ and the total anomalous density $\tilde m=\sum_{\pm}\tilde m_{\pm}$ 
as a function of the small parameter of the theory:
\begin{align}\label{Dep2} 
\frac{\tilde n}{n} &\simeq  \frac{8}{3} \sqrt{ \frac{ n a^3}{\pi} } \sum_{\pm}\left (\frac{\delta g_{\pm}}{g}\right)^{3/2} \bigg\{ \left [1 +\frac{3 \pi} {4} \frac{R'} {(\delta g_{\pm}/g )^2}\right] \\
&-\frac{3}{2} \left[1+\frac{R'} { (\delta g_{\pm}/g )^2} \right]  \frac{v}{c_s \sqrt{\delta g_{\pm}/g}} \bigg\},  \nonumber
\end{align}
and
\begin{align}\label{mDep2} 
\frac{\tilde m}{n} &\simeq  8\sqrt{ \frac{ n a^3}{\pi} }\sum_{\pm}  \left (\frac{\delta g_{\pm}}{g}\right)^{3/2} \bigg\{ \left [1 +\frac{\pi} {4} \frac{R'} {(\delta g_{\pm}/g )^2}\right] \\
&-\frac{1}{2} \left[1+\frac{R'} { (\delta g_{\pm}/g )^2} \right]  \frac{v}{c_s \sqrt{\delta g_{\pm}/g}} \bigg\},  \nonumber
\end{align}
where $R'=R_0 n/m^2 c_s^4$ is a dimensionless disorder strength for each component.
Obviously, for $R'=v=0$, $\tilde n$ and $\tilde m$ reduce to those obtained for symmetric cleaned immovable Bose mixtures \cite{Boudj,Boudj2}.

Substituting Eqs.(\ref{Dep2}) and (\ref{mDep2}) into Eq.(\ref{chim}), the corrected EoS for Bose mixtures with moving components reads
\begin{align}\label{chim1} 
\frac{\mu_{\text{LHY}} }{mc_s^2}&=\frac{32}{3} \sqrt{ \frac{ n a^3}{\pi} }  \sum_{\pm}\left (\frac{\delta g_{\pm}}{g}\right)^{3/2} 
\bigg\{ \left[1 +\frac{3 \pi} {8} \frac{R'} {(\delta g_{\pm}/g )^2}\right] \nonumber\\
&-\left[1+\frac{R'} { (\delta g_{\pm}/g )^2} \right]  \frac{v}{c_s \sqrt{\delta g_{\pm}/g}} \bigg\}. 
\end{align}
For $v = R'= 0$, we reproduce the standard LHY EoS for cleaned Bose mixtures \cite{Boudj2}.
Evidently, for $v = 0$, Eq.(\ref{chim1}) reduces to that of a dirty immovable mixture \cite{BB}.
Again corrections due to the relative motion effects may lead to decrease the chemical potential of the system.
In such a regime the system behaves as a particle-like object with a negative mass 
$M\sim -m\sqrt{ n a^3/\pi }  \sum_{\pm}\left[\left (\delta g_{\pm}/g\right)+R'/ (\delta g_{\pm}/g) \right]$ results in 
any friction force, if exist, would accelerate the mixture.

\begin{figure}
	\includegraphics[width=0.8 \linewidth]{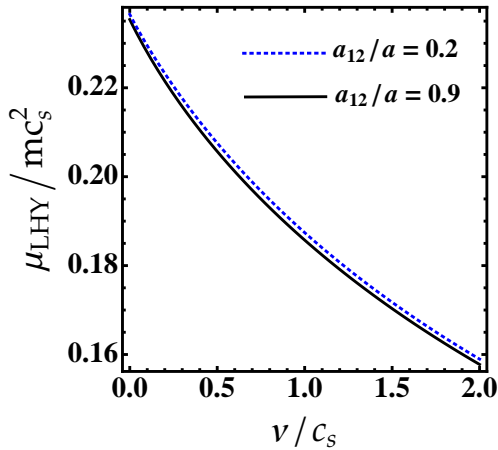}
	\caption{ The LHY corrections to the EoS $\mu_{\text{LHY}} =\sum_{\pm} \mu_{\text{LHY} \pm}$ as a function of  $v/c_s$ for two values of $a_{12}/a$.
Parameters are: $R'=0.05$, and $na^3\sim 10^{-5}$ \cite{Cab}. }
\label {CE}
\end{figure}

The disorder potential can be considered as weak if the condensate depletion due to the disorder is much smaller than the total density $n_R \ll n$. 
This leads to the criterion for the critical disorder strength:
\begin{align} \label{Rcri}
R_c \simeq (n a^3/\pi) ^{-1/2}\sum_{\pm}\frac{\delta g_{\pm}/g} { \sqrt{\delta g_{\pm}/g}-2v/(\pi c_s) }.
\end{align}
For $v=0$, Eq.(\ref{Rcri}) reduces to $R_c \simeq \sum_{\pm} \sqrt{ (\delta g_{\pm}/g)/(n a^3/\pi)}$.
The Bogoliubov theory is no longer applicable for $R' >R'_c$.
Therefore, the investigation of disordered moving binary BECs in such a regime requires to go beyond the perturbative theory or using numerical quantum Monte Carlo simulation.



In order to gain additional insights into the behavior of moving binary BECs in random potentials and to verify our analytical results, 
we perform numerical simulations of the integrals (\ref{Dep})-(\ref{glass}).
Then, we consider two ${}^{39}$K BECs with the interaction parameter of each component is $na^3\sim 10^{-5}$ \cite{Cab},
which is sufficient to ensure that the system meets the requirement of weakly interacting gas.
The interspecies interactions can be adjusted by means of Feshbach resonance.
The disorder strength has been chosen $R'=0.05$. 
However, for such a value the condensate fluctuations due to the disorder is $\lesssim 1\%$ of the total density (see Fig.\ref{AND} (top)).
Evidently, the  disorder fraction $n_R/n$ becomes important for large $R'$.

We observe also that the glassy fraction in both components decreases with $v/c_s$ and increases with $a_{12}/a$.
This indicates that the competition of the relative motion of the mixture fluid, the interspecies interactions and the disorder potential 
may strongly modify the localization of atoms  (see Fig.\ref{AND} (top)).
This is in contrast to the immovable disordered BEC where the disorder tends to fragment the condensate and hence, reduce both the condensed and the superfluid fractions. 
The same holds true for the behavior of the quantum contribution to the noncondensed fraction $\tilde n^0/n$   (see Fig.\ref{AND} (middle)).
The situation is quite different for the anomalous fraction $\tilde m^0/n$, where its modulus decreases with both $v/c_s$ and $a_{12}/a$ (see Fig.\ref{AND} (bottom)).
We are thus treating ultradilute and weakly correlated Bose mixture liquids.

Figure.\ref {CE}  depicts that the LHY corrected EoS is decreasing with $v/c_s$ whatever the value of the ratio $a_{12}/a$ in agreement with our analytical results. 
This is attributed to the interplay of the disorder potential and the relative motion of the mixture.


In conclusion, we systematically investigated the implications of a weak random potential on moving binary BECs with a combined analytical and numerical methods.
Our calculations are based on the Bogoliubov-Huang-Meng approach which is valid for weak interactions, weak disorder and small velocity.
We showed that the relative motion of two components plays a key role in reducing the glassy fraction, the quantum depletion, the anomalous correlations, and the EoS.
One can expect that when the condensate reaches the Landau critical velocity where dissipation is more efficient, the condensate depletion would increase. 
Furthermore, we pointed out that corrections due to the interspecies interaction, disorder and relative motion may lead to the occurrence of stable ultradilute Bose mixtures.
We anticipate that our findings will motivate new avenues of research in quantum transport for disordered Bose mixtures.

We would like to thank V.I. Yukalov for careful reading of the manuscript.

\end{document}